\documentclass[aps,prl,twocolumn,superscriptaddress]{revtex4-1}
\usepackage{amsmath,amssymb,amsfonts}
\usepackage{CJK}
\usepackage{bm}
\usepackage{color}
\usepackage[colorlinks=true, letterpaper=true, pdfstartview=FitV, linkcolor=blue, citecolor=blue, urlcolor=blue]{hyperref}
\usepackage{changes}
\usepackage{appendix}
\usepackage{tikz,pgf}
\usepackage{pdfpages}

\def \H {\mathcal{H}}
\def \E {\mathcal{E}}
\def \L {\mathcal{L}}

\def \bM {\mathsf{M}}
\def \bL {\mathsf{L}}

\def \Z {\mathbb{Z}}

\def \k {\bm{k}}
\def \G {\bm{G}}

\makeatletter
\AtBeginDocument{\let\LS@rot\@undefined}
\makeatother

\makeatletter 
\renewcommand\NAT@biblabelnum[1]{#1.} 
\makeatother

\setcitestyle{super}

\begin{document}

\title{Brillouin Klein Bottle From Artificial Gauge Fields}

\author{Z. Y. Chen}
\affiliation{National Laboratory of Solid State Microstructures and Department of Physics, Nanjing University, Nanjing 210093, China}

\author{Shengyuan A. Yang}
\affiliation{Research Laboratory for Quantum Materials, Singapore University of Technology and Design, Singapore 487372, Singapore}

\author{Y. X. Zhao}
\email[]{zhaoyx@nju.edu.cn}
\affiliation{National Laboratory of Solid State Microstructures and Department of Physics, Nanjing University, Nanjing 210093, China}
\affiliation{Collaborative Innovation Center of Advanced Microstructures, Nanjing University, Nanjing 210093, China}

\begin{abstract}
	A Brillouin zone is the unit for the momentum space of a crystal.  It is topologically a torus, and  distinguishing whether a set of wave
	functions over the Brillouin torus can be smoothly deformed to another leads to the classification of various topological states of matter. Here, we show that under $\Z_2$ gauge fields, i.e., hopping amplitudes with phases $\pm 1$, the fundamental domain of momentum space can assume the topology of a Klein bottle. This drastic change of the Brillouin zone theory is due to the projective symmetry algebra enforced by the gauge field. Remarkably, the non-orientability of the Brillouin Klein bottle corresponds to the topological classification by a $\Z_2$ invariant, in contrast to the Chern number valued in $\Z$ for the usual Brillouin torus. The result is a novel Klein-bottle insulator featuring topological modes at two edges related by a nonlocal twist,
	radically distinct from all previous topological insulators. Our prediction can be readily achieved in various artificial crystals, and the discovery opens a new direction to explore topological physics by gauge-field-modified fundamental structures of physics.
\end{abstract}

\maketitle

\bigskip
	
\noindent \textbf{INTRODUCTION}\\
The Brillouin zone is a fundamental concept in physics. It is essential for the physical description of crystalline solids, metamaterials, and artificial periodic systems. Particularly, it sets the stage for classifying topological states,
which, in mathematical terms, is the task to study the topology of Hermitian vector bundles over the Brillouin zone as the base manifold ~\cite{TKNN,Simon_1983,Schnyder_RMP}.
Clearly, the topology of the Brillouin zone itself is a crucial ingredient for the classification. Since Brillouin zones have the topology of a torus, topological states known to date basically correspond to classifications done on the torus.

Meanwhile, although initially studied for electronic systems in solids~\cite{Volovik_book,Qi:RMP,Kane:RMP}, topological states have been successfully extended to artificial crystals, such as acoustic/photonic crystals, electric circuit arrays, and mechanical networks. These systems have the advantage of great tunability. More importantly, gauge fields can be flexibly engineered in artificial crystals. In particular, the $\mathbb{Z}_2$ gauge field , i.e., hopping amplitudes allowed to take phases $\pm 1$, can be readily realized in these systems and have already been demonstrated in many experiments ~\cite{Ozawa2019rmp,MaGuancong2019nrp,Lu2014,Yang2015,Acoustic_Crystal,Imhof2018,Yu2020,Prodan_Spring,Huber2016,Cooper2019,Optical_Lattice_RMP}. A crucial but so far less appreciated point is that under gauge fields, symmetries of the system would satisfy projective algebras ~\cite{ Wen-PSG,Zhao_PRB_2020,Zhao2021,Shao2021} beyond the textbook group theory for crystal symmetry~\cite{bradley2009mathematical}, which has recently been experimentally demonstrated by acoustic crystals~\cite{Xue2021,Li2021}. Then, what is the physical consequence of the projective symmetry algebra? Does it generate any new topology that is impossible for systems without gauge field? These questions have not been answered yet.

In this article, we reveal that the projective symmetry algebra can lead to a fundamental change of the Bloch
band theory. We show that it can generate a peculiar ``momentum-space nonsymmorphic symmetry", i.e., when represented in momentum space, the projective algebra requires that certain symmetry must include a {fractional} translation in the reciprocal lattice. For example, a real-space reflection symmetry can become a glide reflection in momentum space. This unique feature in turn dictates the topology of the fundamental domain of the momentum space being a Klein bottle and leads to new topological states.

\begin{figure}[ht]
	\includegraphics[width=8.5cm]{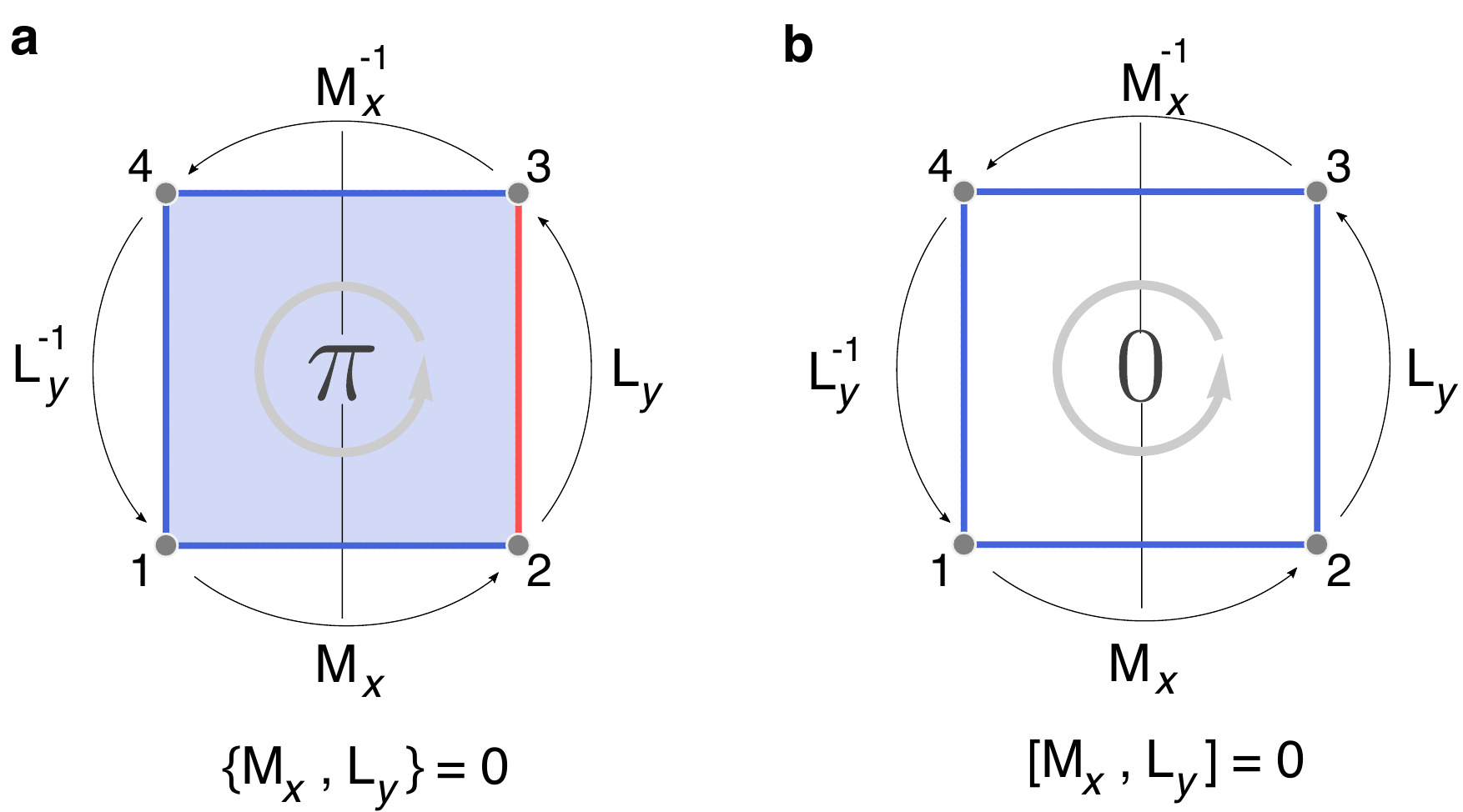}
	\caption{\textbf{Flux and symmetry algebra.} \textbf{a} A rectangle with $\pi$ flux. Blue/red color indicates hopping amplitudes with positive/negative signs.
		Successive operations $\bL_y^{-1}\bM_x^{-1}\bL_y \bM_x$ move a particle around the rectangle, which encloses the $\pi$ flux, so the result is equal to $-1$, leading to the anti-commutation algebra. \textbf{b} When there is no flux, $\bM_x$ and $\bL_y$ follow the ordinary commutation algebra.}
	\label{fig:Momentum_Unit}
\end{figure}

\bigskip
	
\noindent \textbf{RESULTS}\\
\noindent \textbf{Emerged momentum-space glide reflection}
Let us start by considering the reflection symmetry $M_x$ that inverses the $x$ axis, and the translation symmetry $L_y$ along the $y$ direction. In the absence of gauge fields, they should commute with each other $[M_x,L_y]=0$. However, under certain gauge flux configurations, the algebraic relation may be projectively modified to
\begin{equation}\label{Projective_ML}
		\{\bM_x,\bL_y\}=0,
\end{equation}
where the font has been changed to indicate the representations under gauge fields. The seemingly peculiar relation in \eqref{Projective_ML} can be intuitively understood by inspecting Fig.~\ref{fig:Momentum_Unit}. Here, we have four lattice sites forming a rectangle invariant under $M_x$. Assuming there is a $\mathbb{Z}_2$ gauge flux of $\pi$ through the rectangle, then both $\bM_x\bL_y$ and $\bL_y\bM_x$ would send a particle from site 1 to 3, but the two paths encloses a $\pi$ flux, therefore resulting in the anti-commutation.

\begin{figure}[ht]
	\includegraphics[width=8.5cm]{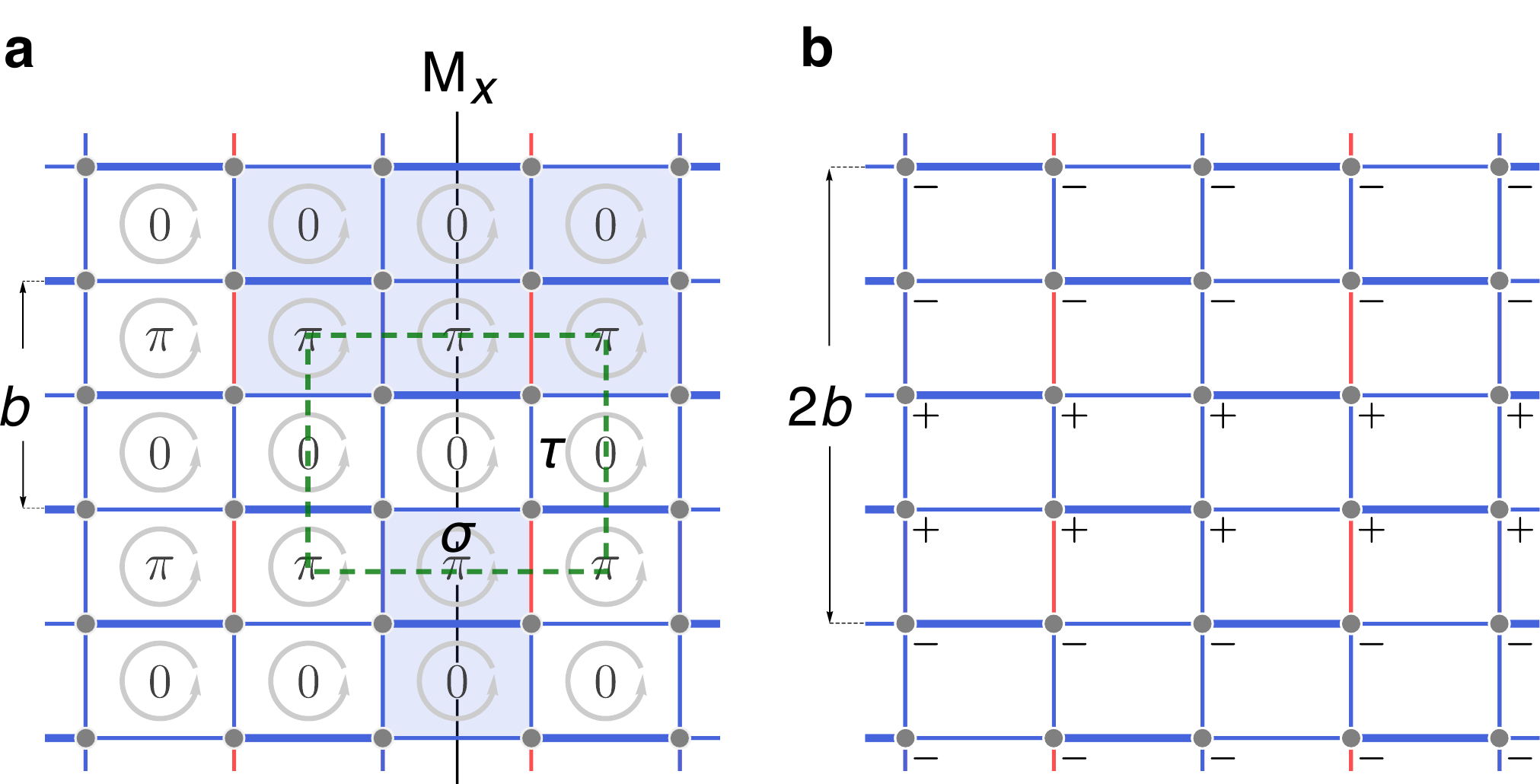}
	\caption{\textbf{A lattice model with non-symmorphic symmetry in the  momentum space.} \textbf{a} The flux configuration and gauge connections of a lattice model. The chosen unit cell is specified by the green dashed rectangle. The blue shaded regions respect $M_x$ and have a unit lattice length along the $y$-direction. Both of them have a net flux $\pi\mod 2\pi$. \textbf{b} The gauge transformation to restore the original gauge connections after reflection $M_x$. }
	\label{fig:fluxed_lattice}
\end{figure}

For a crystalline system, if we choose a unit cell with lattice constant $b$ along the $y$ direction, the operator $\bL_y$ is diagonalized as $\hat{\bL}_y=e^{ik_yb}$ in the momentum space. Then, the projective algebra in Eq.\eqref{Projective_ML} requires
\begin{equation}\label{Half-trans}
	\hat{\bM}_x e^{ik_yb}\hat{\bM}_x=-e^{ik_yb}=e^{i(k_y+G_y/2)b},
\end{equation}
where $G_y$ is the length of the reciprocal lattice vector $\G_y$.
From Eq.~\eqref{Half-trans}, we make the key observation that $\hat{\bM}_x$ must contain a half translation in the reciprocal lattice along $k_y$, when represented in the momentum space. Explicitly,
\begin{equation}\label{Mirror}
		\hat{\bM}_x=U \L_{\frac{\G_y}{2}}\hat{m}_x,
\end{equation}
where $U$ is some unitary matrix,  $\hat{m}_x$ is the operator that inverses $k_x$, and $\L_{{\G_y}/{2}}$ denotes the operator that implements the half translation ${\G_y}/{2}$ of the reciprocal lattice. Hence, $\bM_x$ may be regarded as a momentum-space glide reflection.

As an example, consider the simple lattice model in Fig.~\ref{fig:fluxed_lattice}a. Here, the primitive unit cell in real space consist of four sites.  The $\Z_2$ gauge flux through each plaquette is specified in the figure, respecting $M_x$ and the translation period of $b$ along $y$. Evidently, relation Eq.\eqref{Projective_ML} is fulfilled for this case, and the mirror symmetry operator is represented by
\begin{equation}
	\hat{\bM}_x=\tau_0\otimes\sigma_1 \L_{\frac{\G_y}{2}}\hat{m}_x
\end{equation}
in momentum space, where $\tau$'s and $\sigma$'s are two sets of Pauli matrices that operate on rows and columns of a unit cell [see Fig.~\ref{fig:fluxed_lattice}a].

The appearance of the fractional reciprocal lattice translation can also be understood from the following analysis. To describe a lattice with gauge flux, we need to choose explicit gauge connections on the lattice bonds. For instance, in Fig.~\ref{fig:fluxed_lattice}a, we show a specific gauge choice, with red and blue colors denoting negative and positive hopping amplitudes, respectively. Then, for this given gauge choice, a crystal symmetry operator is given by $\mathsf{R}=\mathsf{G}R$, namely a combination of the manifest spatial operator $R$ and a gauge transformation $\mathsf{G}$. This is because although the flux configuration is invariant under $R$, the specific gauge connection configuration may be changed by $R$. To restore the original gauge connection, an additional gauge transformation $\mathsf{G}$ should be performed. For instance, the gauge transformation required after reflection $M_x$ is depicted in Fig.~\ref{fig:fluxed_lattice}b. Notably, $\mathsf{G}$ may not be compatible with the spatial period of the lattice. [Here, it must be incompatible with $L_y$ due to Eq.\eqref{Projective_ML}.] Clearly, in Fig.~\ref{fig:fluxed_lattice}b, the period of $\mathsf{G}$ along $y$ doubles the lattice constant. Then, after Fourier transform, the incompatibility manifests in $\bM_x$ as a fractional translation in momentum space.
	
\begin{figure}[ht]
	\includegraphics[width=8.5cm]{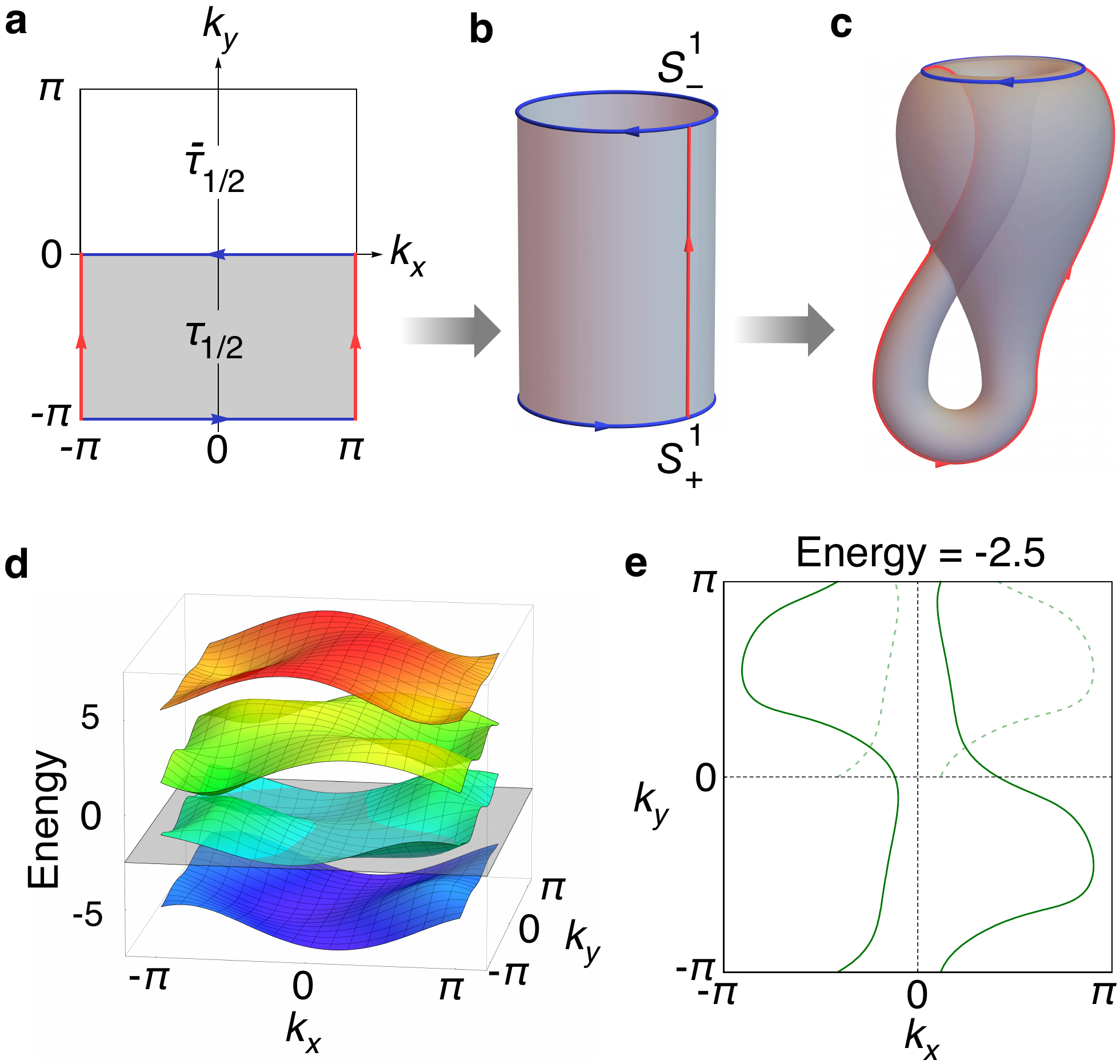}
	\caption{ \textbf{Momentum space representation of the symmetry  and Brillouin Klein-bottle.} \textbf{a} The fundamental domain of the Brillouin zone is $\tau_{1/2}$. The boundaries with the same color should be identified along the marked direction. \textbf{b} The cylinder with two boundaries $S^1_{\pm}$ identified along opposite directions, which is essentially the Brillouin Klein bottle in \textbf{c.} \textbf{d} Energy bands of the model in Fig.~2a. \textbf{e} A constant energy cut, which corresponds to the gray colored plane in \textbf{d}. The reflection of the band structure over $\tau_{1/2}$ through the $k_y$ axis coincides with that over $\bar{\tau}_{1/2}$ after a half translation $\mathcal{L}_{\G_y/2}$. For comparison, in \textbf{e} the curves within $\tau_{1/2}$ are translated to $\bar{\tau}_{1/2}$, and marked as light dashed green lines. }
	\label{fig:Klein_Bottle}
\end{figure}
	
Note that for conventional space groups, nonsymmorphic symmetries such as glide reflections exist only on real-space lattices, i.e., the involved fractional translations act only in real space but not in momentum space~\cite{Wieder_PRB_2016,Watanabe14551}. When transformed to momentum space, they invariably become fixed-point operations, namely, there are always momenta (such as the $\Gamma$ point) that are invariant under the operation. Therefore, ordinary (real-space) nonsymmorphic symmetries are fundamentally distinct from the
momentum-space nonsymmorphic symmetry $\bM_x$ discovered here, for which the fractional translation acts in momentum space. It is also clear that
$\bM_x$ is a \emph{free} operation, i.e., no momentum is invariant under $\bM_x$.  The emergence of momentum-space nonsymmorphic symmetry is a unique feature of projective symmetry algebras. The free character of such symmetry operations will produce remarkable consequences, as we discuss below.
	
\smallskip
	
\noindent\textbf{Brillouin Klein bottle} We proceed to elucidate the physical consequences of this momentum-space glide reflection symmetry.
Let $\H(\k)$ be the Bloch Hamiltonian in momentum space. Then, the constraint by $\bM_x$ in Eq.\eqref{Mirror} is
\begin{equation}
		U\H(k_x,k_y)U^\dagger=\mathcal{H}(-k_x,k_y+\pi).
\end{equation}
Here, for simplicity we have set $b=1$.
This means if $|\psi(\k)\rangle$ is an eigenstate of $\H(\k)$ with energy $\E(\k)$, then $U|\psi(\k)\rangle$ will be an eigenstate of $\H(-k_x,k_y+\pi)$ with the same energy, i.e.,
\begin{equation}\label{sym-condi}
		\H(-k_x,k_y+\pi)U|\psi(\k)\rangle=\E(\k) U|\psi(\k)\rangle.
\end{equation}
As a result the spectrum at $(k_x,k_y)$ is equivalent to that at $(-k_x,k_y+\pi)$. Thus, the Brillouin zone can be partitioned into two parts, $\tau_{1/2}$ and $\bar{\tau}_{1/2}$, as illustrated in Fig.~\ref{fig:Klein_Bottle}a. Only one of them is independent, i.e., the fundamental domain of momentum space is a half of the Brillouin zone.
	
This can be explicitly verified for the model in Fig.~\ref{fig:fluxed_lattice}. In Fig.~\ref{fig:Klein_Bottle}d, we plot the spectrum of the lattice model. One can observe that the reflection of the band structure over $\tau_{1/2}$ through the $k_y$ axis coincides with that over $\bar{\tau}_{1/2}$ after a half translation $\mathcal{L}_{\G_y/2}$.
This can be more clearly seen from the constant energy cut in Fig.~\ref{fig:Klein_Bottle}e.
	
We have emphasized that as a momentum-space nonsymmorphic symmetry, $\bM_x$ is a \emph{free} operation with no fixed point, distinct from conventional space group symmetries. Mathematically, it is known that an equivariant bundle with the structure group $G$ \emph{freely} acting on the based space $X$ is \emph{equivalent} to the bundle on the orbital space $X/G$ \cite{segal1968equivariant}. For our case, this simply means all the information including topology is fully captured by the fundamental domain
$\tau_{1/2}=[-\pi,\pi)\times[-\pi,0)$. Since $k_x$ is periodic, we may write $\tau_{1/2}=S^1\times [-\pi,0)$ as a cylinder. This cylinder has two boundaries $S^1_{\pm}$ at $k_y=-\pi$ and $0$, respectively. Importantly, $S^1_{\pm}$ are oppositely oriented and ``glued" together, because they are connected by $\bM_x$ [Fig.~\ref{fig:Klein_Bottle}b]. Thus, the fundamental domain here is topologically a Klein bottle, as illustrated in Fig.~\ref{fig:Klein_Bottle}c.
	
We remark that in solid state physics, conventional space group symmetries are commonly used to reduce Brillouin zones to so-called irreducible Brillouin zones. However, because those symmetries are not free, the irreducible Brillouin zone is not sufficient to capture the topological information (e.g., symmetry information is still required at high-symmetry points or paths of the irreducible Brillouin zone), distinct from the case here. Besides, the Brillouin Klein bottle is a closed manifold, whereas the irreducible Brillouin zones are not. These characters are important for the topological classification to be discussed in the following.
	
\smallskip
	
\noindent\textbf{Topological invariant and edge states}
Consider the system is in an insulating phase. The task of topological classification is to classify valence band wave functions (forming a Hermitian vector bundle) over the Brillouin Klein bottle. This is fundamentally different from the usual cases where the base manifold is a torus or a sphere.
A crucial difference is the orientability. A torus (and a sphere) is orientable, whereas a Klein bottle is non-orientable. For orientable closed base manifolds such as the torus, the most elementary topological invariant is the Chern number, which is the integration of the Berry curvature $\mathcal{F}$ for valence bands over the Brillouin torus. The Chern number is valued in $\mathbb{Z}$, and the sign of the integer is related to the orientation of the torus, since a reflection inverses the Chern number. In contrast, for the Brillouin Klein bottle which is non-orientable, any topological invariant can only be valued in $\Z_2$, since the sign of the invariant has no significance and we must have $1=-1$. 

\begin{figure}[ht]
	\includegraphics[width=8.5cm]{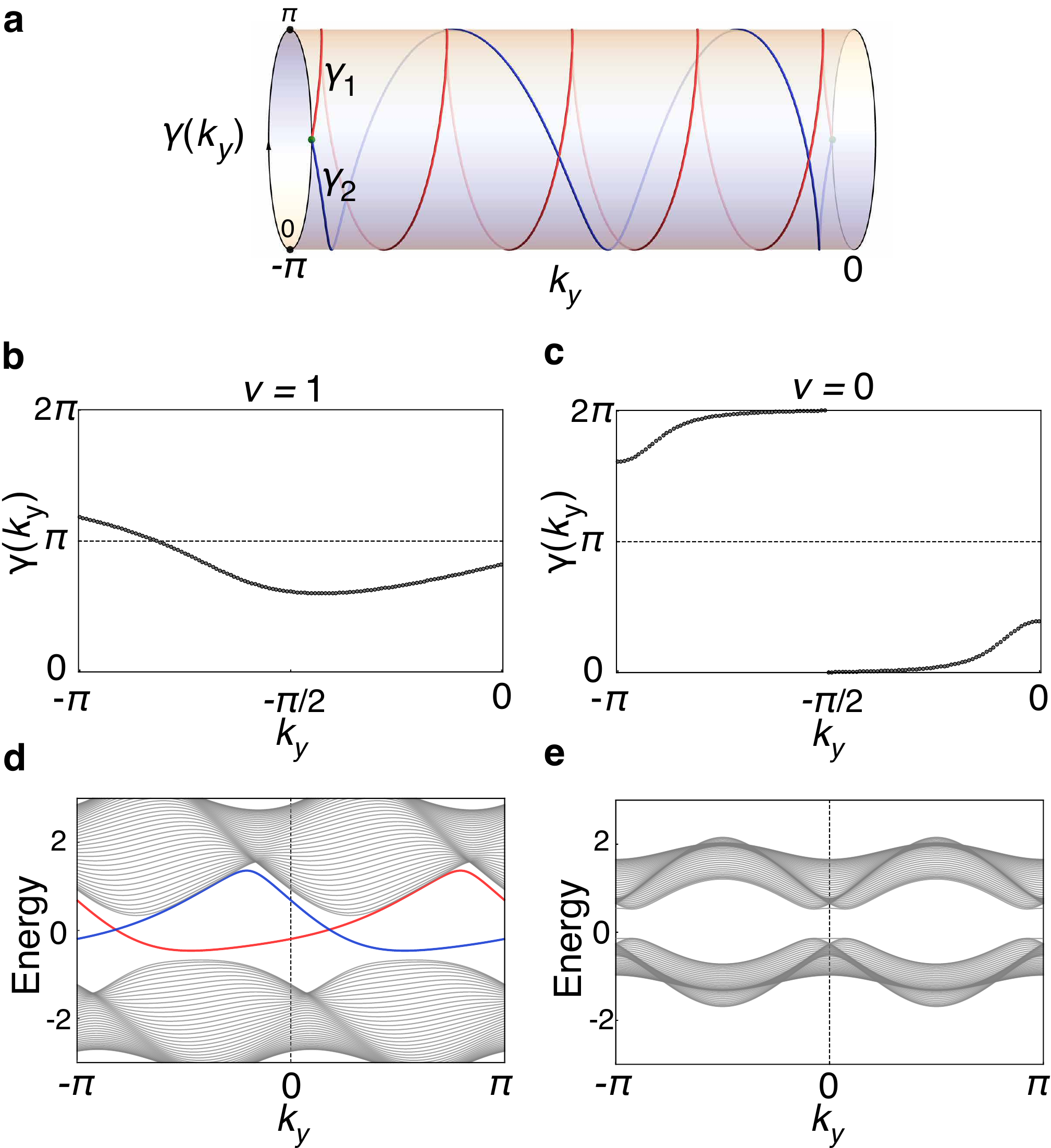}
	\caption{\textbf{Topological invariant and  edge states.} \textbf{a} Schematic illustration of the formula Eq.\eqref{Parity}. The red and blue paths correspond to the topologically nontrivial and trivial cases, respectively. \textbf{b} and \textbf{c} depict the flows of $\gamma(k_y)$ for two set of parameters for the model in Fig.~2a, and the corresponding band structures on a ribbon geometry with edges along $y$ are given in \textbf{d} and \textbf{e}, respectively. States on right/left edge are marked in red/blue color. }
	\label{fig:Top_Invar}
\end{figure}

Now, we formulate an explicit expression for this $\Z_2$ topological invariant. This is based on two key observations. First, the two boundaries $S_{\pm}^1$ of $\tau_{1/2}$ are related by an inversion of $k_x$ [Fig.~\ref{fig:Klein_Bottle}a]. The inversion operation inverses the Berry phase for a $1$D system~\cite{Zak_Phase}. Hence, the Berry phases $\gamma(-\pi)$ and $\gamma(0)$ over $S_{\pm}^1$ are opposite up to an multiple of $2\pi$, i.e., $\gamma(0)+\gamma(-\pi)=0\mod 2\pi$. Second, due to Stoke's theorem, $\int_{\tau_{1/2}}d^2k~\mathcal{F}+\gamma(0)-\gamma(-\pi)=0\mod 2\pi$. Therefore, we can formulate the $\Z_2$ invariant as
\begin{equation}\label{Topo_Invariant}
		\nu=\frac{1}{2\pi}\int_{\tau_{1/2}}d^2k~\mathcal{F}+\frac{1}{\pi}\gamma(0) \mod 2.
\end{equation}
Here, the formula is valued in integers because of the two observations above. Since a large gauge transformation for valence wave functions can change $\gamma(0)$ by a multiple of $2\pi$, only the parity of the formula is gauge invariant and hence can be defined as a topological invariant. We also comment that the $\Z_2$ topological classification here is based on the equivariant K theory or KG theory by $\tilde{\mathrm{K}}(K)=\mathbb{Z}_2$, where $K$ is the Klein bottle. The topological invariant is derived from the fact that line bundles over a $2$D manifold $M$ are topologically classified by $H^2(M,\mathbb{Z})$, with $H^2(K,\mathbb{Z})=\mathbb{Z}_2$. Thus the resultant classification and the topological invariant are stable under the addition of trivial bands. 

We can give Eq.~\eqref{Topo_Invariant} a pump interpretation with an intuitive geometric picture. Over $\tau_{1/2}=S^1\times[-\pi,0]$, we can always choose a complete set of continuous valence states $|\psi_n(\k)\rangle$, which are periodic along $k_x$. Then, the corresponding Berry connection $\mathcal{A}(\k)$ is also periodic in $k_x$. For such a $\mathcal{A}(\k)$, we can compute $\gamma(k_y)$ that is continuous from $k_y=-\pi$ to $0$. Moreover, it is straightforward to derive that $\int_{\tau_{1/2}}d^2k\mathcal{F}=-\int_{-\pi}^0dk_y~\partial_{k_y}\gamma(k_y)=\gamma(-\pi)-\gamma(0)$. Hence, from Eq.~\eqref{Topo_Invariant}, we find that
\begin{equation}\label{Winding}
\nu=\frac{1}{2\pi}[\gamma(0)+\gamma(-\pi)] \mod 2.
\end{equation}
Considering the generic case that $\gamma(-\pi)\ne 0$ or $\pi$, the path of $\gamma(k_y)$ has to cross $0$ or $\pi$ in the course of varying $k_y$ from $-\pi$ to $0$ [see Fig.~\ref{fig:Top_Invar}a].
Introducing $W_{0\slash\pi}$ as the number of times that $\gamma(k_y)$ crosses $0$/$\pi$, $\nu$ can be given a geometric interpretation:
\begin{equation}\label{Parity}
	\nu=W_\pi \mod 2,
\end{equation}
i.e., $\nu$ is nontrivial if and only if $\gamma(k_y)$ crosses $\pi$ an odd number of times.
	
The insulator with nontrivial $\nu=1$ may be termed as a Klein-bottle insulator. It features special topological edge states, whose existence can be understood from Eq.\eqref{Parity}.  For nontrivial $\nu$, $\gamma(k_y)$ has to cross $\pi$ at some (odd number of) $k_y$, then the $1$D $k_x$-subsystems at these crossing points have Berry phases of $\pi$. It is well known that the valence-band Berry phase correspond to the center of Wannier function and the 1D charge polarization. Particularly, $\gamma=\pi$ corresponds to Wannier center at the midpoints between lattice sites, and therefore leads to an in-gap state at each end. Thus, there must be in-gap boundary states located at each edge parallel to the $y$ direction. Because of the continuity of energy bands, these in-gap states must be connected to form a topological edge band.

Consider our model in Fig.~\ref{fig:fluxed_lattice} with two sets of parameters. The topological invariant Eq.\eqref{Parity} is computed as shown in Fig.~\ref{fig:Top_Invar}b and c, respectively. The corresponding band structures for a ribbon geometry with edges along the $y$ direction.
are shown in Fig.~\ref{fig:Top_Invar}d and e. The topological edge bands are clearly observed for the Klein-bottle insulator phase with $\nu=1$. 
Similar to the M\"{o}bius insulators, these edge bands are detached from the bulk bands~\cite{Sato_Mobius,Zhao_PRB_2020,Young_Wieder_PRL_2017}. Under strong boundary potentials, they could be shifted out of the gap.

We note two features of the edge states that distinguish the Klein bottle insulator from conventional crystalline topological insulators. First, the gapless modes appear on mirror-symmetry-breaking edges, rather than on symmetry-preserving edges as for conventional crystalline topological insulators. It is easy to see that the above argument indicates the existence of topological edge modes on any edge {not} perpendicular to $y$, whereas the mirror-symmetric edge perpendicular to $y$ is expected to be gapped without topological edge mode. Second, the momentum-space glide reflection fascinatingly leads to a nonlocal relation for edge states.
Consider two edges along $y$ connected by the $M_x$ symmetry in real space,
Because of the nonsymmorphic character of ${\mathsf{M}}_x$ in the momentum space, only the energy bands over $k_y\in[-\pi,0)$ are independent, while those over $k_y\in[0,\pi)$ can be deduced from the action of ${\mathsf{M}}_x$.  Particularly, ${\mathsf{M}}_x$ nonlocally maps the topological edge band on one edge over $k_y\in[-\pi,0)$ to that on the other edge over $k_y\in[0,\pi)$. In Fig.~\ref{fig:Top_Invar}d, one can clearly see that translating the edge band on the left edge by $G_y/2$ coincides with that on the right edge.

We have demonstrated that interplay between gauge fields and symmetry can fundamentally modify the Bloch band theory. Under gauge fields, a spatial symmetry can acquire a nonsymmorphic character in momentum space. Particularly, the momentum-space glide reflection can reduce the Brillouin torus to the Brillouin Klein bottle,  and therefore change the topological classifications from the bottom level. We formulate a novel kind of topological insulator over the Brillouin Klein bottle, which is as elementary as the Chern insulator over the Brillouin torus. Although we take 2D reflection in our analysis, the discussion can be readily generalized to the $3$D with analogous momentum-space glide reflections and screw rotations  (see Supplementary Note 4 for demonstrations). Since glide reflection and screw rotations are the most elementary nonsymmorphic symmetries, all nonsymmorphic space groups may be realized on the reciprocal lattices by certain gauge flux configurations, which are mathematically dictated by the second cohomology groups of the space groups. Since gauge fluxes can be engineered in artificial crystals for realizing projective symmetries~\cite{Xue2021,Li2021}, 
our work opens the door towards a fertile ground for exploring novel momentum-space symmetries and topologies of artificial crystals beyond the scope of topological quantum materials.

\bigskip

\noindent \textbf{METHODS}\\
\textbf{The simple 2D model.} We consider a model defined on the rectangular lattice in Fig.~\ref{fig:fluxed_lattice}. Constrained by $M_x$ and two translation symmetries, the most general Hamiltonian with only nearest neighbor hopping terms is given by
\begin{equation}\label{Model}
		\H_0(\k)=	\begin{bmatrix}
			\varepsilon & [q^x_1(k_x)]^*  & [q^y_{+}(k_y)]^*  & 0 \\
			q^x_1(k_x)  & \varepsilon & 0 & [q^y_{-}(k_y)]^*  \\
			q^y_+(k_y)  & 0 & -\varepsilon & [q_2^x(k_x)]^* \\
			0 & q^y_{-}(k_y)  & q_2^x(k_x)   & -\varepsilon\\
		\end{bmatrix},
\end{equation}
where  $q^x_{a}(k_x)=t_{a1}^x+t_{a2}^x e^{ik_x}$ with $a=1,2$, $q^y_{\pm}(k_y)=t_{1}^y\pm t_{2}^y e^{ik_y}$, $\pm\varepsilon$ are on-site energies. To break the time-reversal ($T$) symmetry, we may include the following second neighbor hopping terms, $\H^{(1)}(\k)=\lambda \cos k_y \tau_1\otimes\sigma_2+\lambda \sin k_y\tau_2\otimes\sigma_2$. For Fig.\ref{fig:Klein_Bottle}de and Fig.\ref{fig:Top_Invar}bd, the parameter are given by $t^x_{11} = t^x_{22} = 1,t^x_{12} = t^x_{21} = 3.5,t^y_1 = 2,t^y_2 = 1.5, \varepsilon = 1,\lambda=1$. For Fig.\ref{fig:Top_Invar}ce, $t^x_{11} = t^x_{12} = 1,t^x_{21} = 3.5, t^x_{22} = 1.7,t^y_1 = 2,t^y_2 = 1.5, \varepsilon = 0.6,\lambda=0$.
	
It is worth pointing out that if the $T$ symmetry is preserved, the two $1$D $k_x$-subsystems $\H(k_x,\pm\pi/2)$ are invariant under $\bM_xT$. This is because $T$ inverses $(k_x,\pm \pi/2)$ to $(-k_x,\mp\pi/2)$, but $\bM_x$ moves $(-k_x,\mp\pi/2)$ back to $(k_x,\pm\pi/2)$. Then, $\bM_xT$ is effectively a spacetime inversion symmetry for $\H(k_x,\pm\pi/2)$, and therefore can quantize its Berry phases into integral multiples of $\pi$. As a result, the curve in Fig.\ref{fig:Top_Invar}b would always cross $\pi$ at $k_y=-\pi/2$.
	
\bigskip
\noindent \textbf{DATA AVAILABILITY}\\
The data generated and analyzed during this study are available from the corresponding author upon reasonable request.

\bigskip
	
\def\bibsection{\ } 
\noindent \textbf{REFERENCES}
\bibliographystyle{naturemag}
\bibliography{Kleinbottle}

\bigskip
	
\noindent \textbf{ACKNOWLEDGEMENTS}\\
This work is supported by  National Natural Science Foundation of China (Grants No. 11874201 and No. 12174181), and the Singapore MOE AcRF Tier 2 (MOE2019-T2-1-001).
		
\bigskip

\noindent \textbf{AUTHOR CONTRIBUTIONS}\\
Z.C. and Y.Z. conceived the idea. S.Y. and X.Z. supervised the project. Z.C. and Y.Z. did the theoretical analysis.
Z.C., S.Y. and Y.Z. wrote the manuscript.

\bigskip
\noindent \textbf{COMPETING INTERESTS}\\
The authors declare no competing interests.

\widetext
\clearpage


\section*{Supplementary material}

# Supplemental Information for
# "Brillouin Klein Bottle From Artificial Gauge Fields"

**Contents**

**Supplementary Note 1. Crystal Symmetries with Gauge Fields**

First of all, we present a formalism of crystal symmetries on a lattice with gauge fields. Only gauge flux configurations are gauge invariant. A given gauge flux configuration can be described by numerous corresponding gauge connection configurations. Reversely, two gauge connection configurations corresponding to the same flux configuration are related by a gauge transformation. A gauge transformation $\mathsf{G}$ simply multiplies a phase for each lattice site, and therefore can be regarded as a diagonal matrix indexed by the lattice sites.

The spatial symmetries are determined by the flux configuration, which is described by a chosen gauge connection configuration. A spatial symmetry $R$ that preserves the flux configuration in general changes the connection configuration. The transformed connection configuration must be related to the original one by a gauge transformation $\mathsf{G}_R$, since they correspond to the same flux configuration. Thus, the physical operator for the spatial symmetry $R$ is given by

$$\mathsf{R} = \mathsf{G}_R R, \tag{1}$$

a combination of spatial symmetry $R$ and the gauge transformation $\mathsf{G}_R$.

We now consider how the physical operator $\mathsf{R}$ acts on a tight-binding model, $H = \sum_{ij} t_{ij} |i\rangle\langle j|$. The transformation of $H$ is given by

$$H' = \sum_{ij} t_{ij} |\mathsf{G}_R R(i)\rangle\langle \mathsf{G}_R R(j)| \tag{2}$$

$$= \sum_{ij} t_{ij} \mathsf{G}_R(R(i)) \mathsf{G}_R^*(R(j)) |R(i)\rangle\langle R(j)|$$

$$= \sum_{ij} \mathsf{G}_R(i) \mathsf{G}_R^*(j) t_{R^{-1}(i)R^{-1}(j)} |i\rangle\langle j|, \tag{3}$$

where $G_R(i)$ is the phase for site $i$, and $R(i)$ is the site transformed from $i$ by $R$. Thus, the invariance under $\mathsf{R}$ gives

$$t_{ij} = \mathsf{G}_R(i) \mathsf{G}_R^*(j) t_{R^{-1}(i)R^{-1}(j)}. \tag{4}$$

**Supplementary Note 2. The Flux Requirement for Momentum-space Glide Reflection Symmetry**

In this section, we show that the projective algebraic relation, Eq. (1) in the main text, leads to the requirement on the flux configuration as mentioned in the main text.

Supplementary Eq. (1) leads to

$$\mathsf{M}_x \mathsf{L}_y \mathsf{M}_x^{-1} \mathsf{L}_y^{-1} = \mathsf{G}_M M_x \mathsf{G}_y L_y (\mathsf{G}_M M_x)^{-1} (\mathsf{G}_y L_y)^{-1} \tag{5}$$

$$= \mathsf{G}_M (M_x \mathsf{G}_y M_x^{-1}) M_x L_y M_x^{-1} L_y^{-1} (L_y \mathsf{G}_M L_y^{-1}) \mathsf{G}_y^{-1}$$

$$= \mathsf{G}_M(\boldsymbol{r}) \mathsf{G}_y(M_x(\boldsymbol{r})) \mathsf{G}_M^*(L_y^{-1}(\boldsymbol{r})) \mathsf{G}_y^*(\boldsymbol{r}). \tag{6}$$

Hence, the projective algebraic relation gives the identity,

$$\mathsf{G}_M(\boldsymbol{r}) \mathsf{G}_y(M_x(\boldsymbol{r})) \mathsf{G}_M^*(L_y^{-1}(\boldsymbol{r})) \mathsf{G}_y^*(\boldsymbol{r}) = -1. \tag{7}$$

Consider a rectangle invariant under $M_x$ and have a unit lattice length along the $y$-direction as Supplementary Fig.1a. Symmetry $\mathsf{M}_x$ and $\mathsf{L}_y$ constrain the hopping amplitudes by

$$t_{24} = \mathsf{G}_M(2)\mathsf{G}_M^*(4) t_{13}, \quad t_{34} = \mathsf{G}_y(3)\mathsf{G}_y^*(4) t_{12}. \tag{8}$$

Then, the flux through the rectangle satisfies

$$e^{-i\Phi} = e^{i\phi_{12}} e^{i\phi_{24}} e^{i\phi_{43}} e^{i\phi_{31}} \tag{9}$$

$$= \mathsf{G}_M(2)\mathsf{G}_M^*(4)\mathsf{G}_y^*(3)\mathsf{G}_y(4) e^{i\phi_{12}} e^{i\phi_{13}} e^{i\phi_{21}} e^{i\phi_{31}}$$

$$= \mathsf{G}_M(2)\mathsf{G}_M^*(4)\mathsf{G}_y^*(3)\mathsf{G}_y(4)$$

$$= \mathsf{G}_M(L_y^{-1}(4))\mathsf{G}_M^*(4)\mathsf{G}_y^*(M_x(4))\mathsf{G}_y(4) = -1. \tag{10}$$

Thus, we have proved the flux requirement, i.e., the magnetic flux through this rectangle is $\pi$.

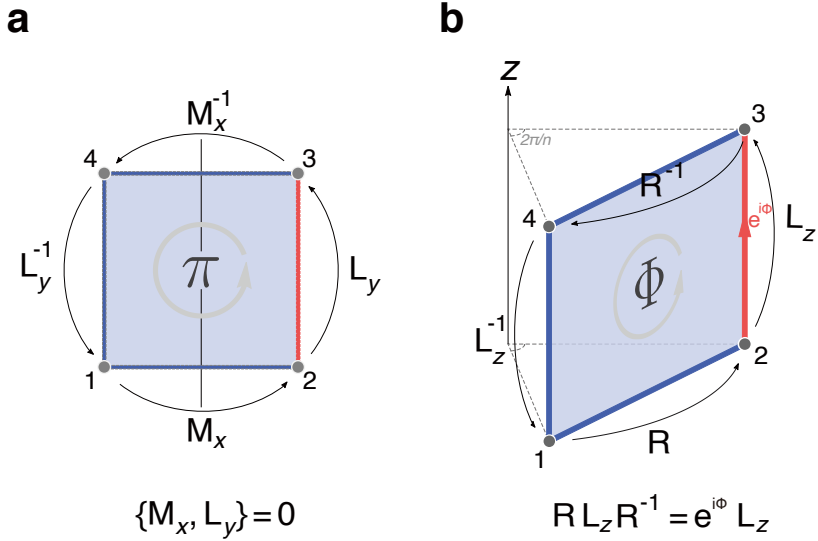

Supplementary Fig.1. **Flux configurations for the projective algebraic relations**. **a** The flux through a rectangle with unit lattice length along the $y$ direction and invariant under the reflection. We choose the gauge condition: All blue bonds have a hopping phases $+1$, and red bond has a hopping phase $-1$. **b** Flux through a rectangle spanned by translation and the rotation. We choose the gauge condition: All blue bonds have a hopping phase 1, and the red bond has a hopping phase $e^{i\Phi}$.

### Supplementary Note 3. Detailed derivation for the glide reflection operator

We now derive the glide reflection operator of the model in the main text. We consider the gauge condition with $\mathsf{G}_y = 1$. Then, Supplementary Eq.(7) becomes $\mathsf{G}_M(\boldsymbol{R})\mathsf{G}_M^*(L_y^{-1}(\boldsymbol{R})) = -1$, which means the period of gauge transformation is half of the lattice period along the $y$ direction. We label the lattice sites by $(\alpha, \boldsymbol{R})$, where $\alpha$ labels sites within a unit cell and $\boldsymbol{R}$ coordinates unit cells. Because the gauge transformation $\boldsymbol{G}_M$ alternates its sign in the $y$ direction, i.e., $\mathsf{G}_M(\alpha, \boldsymbol{R}) = -\mathsf{G}_M(\alpha, \boldsymbol{R} + \boldsymbol{e}_y)$, it transforms local states $|\alpha, \boldsymbol{R}\rangle$ as

$$|\alpha, \boldsymbol{R}\rangle \xrightarrow{\mathsf{G}_M} (-1)^{n_y} \mathsf{G}_{M\alpha} |\alpha, \boldsymbol{R}\rangle = e^{i\pi n_y} \mathsf{G}_{M\alpha} |\alpha, \boldsymbol{R}\rangle \tag{11}$$

$$= e^{i\boldsymbol{G}_y \cdot \boldsymbol{R}/2} \mathsf{G}_{M\alpha} |\alpha, \boldsymbol{R}\rangle,$$

where $\mathsf{G}_{M\alpha} = \mathsf{G}_M(\alpha, \mathbf{0})$ is the gauge transformation within a unit cell. The momentum-space states $|\alpha, \mathbf{k}\rangle$ are the Fourier transform of the real-space states, i.e., $|\alpha, \mathbf{k}\rangle = \sum_{\mathbf{R}} e^{i\mathbf{k}\cdot\mathbf{R}}|\alpha, \mathbf{R}\rangle$. Thus, $\mathsf{G}_M$ transforms the momentum-space states as

$$|\alpha, \mathbf{k}\rangle \xrightarrow{\mathsf{G}_M} \mathsf{G}_{M\alpha}|\alpha, \mathbf{k} + \mathbf{G}_y/2\rangle. \tag{12}$$

Hence, its action on the momentum space contains a half translation operator. Combining the reflection operator with the gauge transformation, the proper reflection in the momentum space is an glide-reflection,

$$\hat{\mathsf{M}}_x = U\mathcal{L}_{\frac{\mathbf{G}_y}{2}}\hat{m}_x. \tag{13}$$

For our model in the main context, in the absence of gauge fields, the momentum-space refection operator is $\hat{M}_x = \tau_0 \otimes \sigma_1 \hat{m}_x$. When we have gauge fields, only the signs of the intercell hopping amplitudes in the $y$ direction are changed after reflection. Hence, the gauge transformation is simply given by $\mathsf{G}_M(\alpha, \mathbf{R}) = (-1)^{n_y}$. Accordingly, the momentum space symmetry operator is given by

$$\hat{\mathsf{M}}_x = \mathcal{L}_{\frac{\mathbf{G}_y}{2}}\hat{M}_x = \tau_0 \otimes \sigma_1 \mathcal{L}_{\frac{\mathbf{G}_y}{2}}\hat{m}_x. \tag{14}$$

### Supplementary Note 4. Screw Rotations in Momentum Space

In this section, we show how to realize screw rotations in momentum space. Without loss of generality, let us consider a rotation group $C_n$ of order $n$ in the $xoy$ plane. In the absence of gauge field, the rotation generator $R$ commutes with translation along $z$ direction $RL_zR^{-1} = L_z$. When there are fluxes through the plaquettes spanned by $L_z$ and $R$ as illustrated in Supplementary Fig.1b, this equation is modified with an additional phase factor, i.e.,

$$RL_zR = e^{i\Phi}L_z. \tag{15}$$

The projective algebraic relation leads to

$$\begin{aligned}\mathsf{R}\mathsf{L}_z\mathsf{R}^{-1}\mathsf{L}_z^{-1} &= \mathsf{G}_R R \mathsf{G}_z L_z (\mathsf{G}_R R)^{-1}(\mathsf{G}_z L_z)^{-1} \\ &= \mathsf{G}_R(R\mathsf{G}_z R^{-1})RL_zR^{-1}L_z^{-1}(L_z\mathsf{G}_R L_z^{-1})\mathsf{G}_z^{-1} \\ &= \mathsf{G}_R(\mathbf{r})\mathsf{G}_z(R(\mathbf{r}))\mathsf{G}_R^*(L_z^{-1}(\mathbf{r}))\mathsf{G}_z^*(\mathbf{r}) = e^{i\Phi}.\end{aligned} \tag{16}$$

Consider the plaquette illustrated in Supplementary Fig.1b. According to Supplementary Eq. (4), the hopping amplitudes are related by

$$t_{24} = \mathsf{G}_R(2)\mathsf{G}_R^*(4)t_{13}, \quad t_{34} = \mathsf{G}_z(3)\mathsf{G}_z^*(4)t_{12}. \tag{17}$$

Then, we can substitute the identities into the formula of flux, which leads to some useful identities.

$$\begin{aligned}e^{-i\Phi} &= e^{i\phi_{12}}e^{i\phi_{24}}e^{i\phi_{43}}e^{i\phi_{31}} \\ &= \mathsf{G}_R(2)\mathsf{G}_R^*(4)\mathsf{G}_z^*(3)\mathsf{G}_z(4)e^{i\phi_{12}}e^{i\phi_{13}}e^{i\phi_{21}}e^{i\phi_{31}} \\ &= \mathsf{G}_R(2)\mathsf{G}_R^*(4)\mathsf{G}_z^*(3)\mathsf{G}_z(4) \\ &= \mathsf{G}_R(L_z^{-1}(4))\mathsf{G}_R^*(4)\mathsf{G}_z^*(R(4))\mathsf{G}_z(4).\end{aligned} \tag{18}$$

Because the rotation is n-fold, $R^n$ is an identity operator, which means

$$\mathsf{L}_z = \mathsf{R}^n \mathsf{L}_z \mathsf{R}^{-n} = e^{in\Phi}\mathsf{L}_z. \tag{19}$$

Thus, $\Phi$ must be quantized, and therefore is valued in $\mathbf{Z}_n$, i.e., $\Phi = 2\pi m/n$ where $m$ is an integer.

When $m \neq 0$, this projective algebraic relation can lead to a screw rotation in momentum space. In momentum space, we choose unit length $c$ along $z$ direction specified by $L_z$ and gauge condition satisfies $\mathsf{G}_z = 1$, then the translation operator is diagonalized as $\hat{\mathsf{L}}_z = e^{ik_z c}$. The projective algebraic relation becomes

$$\hat{\mathsf{R}}e^{ik_z c}\hat{\mathsf{R}}^{-1} = e^{in\Phi}e^{ik_z c} = e^{i(k_z c + G_z m/n)}. \tag{20}$$

Thus, the proper transformation of R in momentum space contains a fractional translation on the reciprocal lattice, and R is represented as

$$\hat{\mathsf{R}} = U\mathcal{L}_{\frac{m\boldsymbol{G}_z}{n}}\hat{r}, \quad (21)$$

where $\mathcal{L}_{\frac{m\boldsymbol{G}_z}{n}}$ is the fractional translation on the reciprocal lattice, $\hat{r}$ rotates $\boldsymbol{k}$ vectors, and $U$ is a unitary matrix.

With the screw-rotation in momentum space, the fundamental domain of Brillouin zone is $1/n$ of origin ones. But different from the glide-reflection, the screw rotations in momentum space preserve the orientation of the the Brillouin 3-torus, so the the topology of fundamental domain is still the same as 3-torus.

To illustrate the screw rotations in momentum space, we construct lattice models for $n = 2, 3, 4, 6$, which are, respectively, illustrated in Supplementary Fig.2abcd. It is noteworthy that these are all possible rotation degrees for crystalline systems. We choose an particular gauge condition: Only red bonds take a nontrivial hopping phase. When $n > 2$, let $q$ be the number of arrow heads in each red bond. Then, the hopping amplitude is given by $t_{i,i+e_z} = e^{-iq\Phi}|t_{i,i+e_z}|$. Below, we list the model Hamiltonians and the corresponding symmetry operators in momentum space.

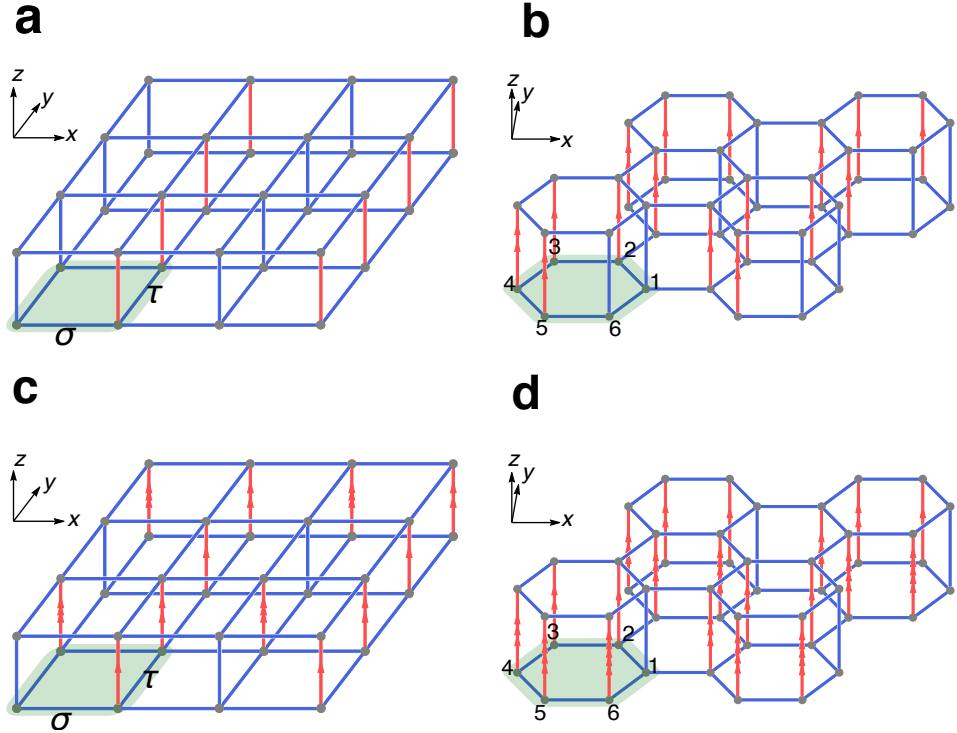

Supplementary Fig.2. **Lattice model with gauge connection satisfies n-folds screw rotation in the momentum space**. **a** $n = 2$. Flux in two nearby vertical surface surrounding the unit cell (green-shaded region) is $\pi$. **b** $n = 3$. Flux in two nearby vertical surface surrounding the unit cell is $\Phi = 2\pi m/3$ and $m = 1, 2$. **c** $n = 4$. Flux in each vertical surface surrounding the unit cell is $\Phi = \pi m/2$ and $m = 1, 2, 3$. **d** $n = 3$. Flux in each vertical surface surrounding the unit cell is $\Phi = \pi m/3$ and $m = 1, 2, 3, 4, 5$.

**a** $n = 2$. The rotation symmetry operator is represented by

$$\hat{\mathsf{R}}_\pi = \tau_1 \otimes \sigma_1 \mathcal{L}_{\frac{m\boldsymbol{G}_z}{2}} \hat{r}_\pi, \quad (22)$$

where $m = 1$, $\tau$'s and $\sigma$'s are two sets of the Pauli matrices that operate on rows and columns of a unit cell. The Hamiltonian is given by

$$\mathcal{H}(\boldsymbol{k}) = \begin{bmatrix} 2t^z \cos k_z & t_1^x + t_2^x e^{-ik_x} & t_1^y + t_2^y e^{-ik_y} & 0 \\ t_1^x + t_2^x e^{ik_x} & 2t'^z \cos(k_z - \Phi) & 0 & t_1^y + t_2^y e^{-ik_y} \\ t_1^y + t_2^y e^{ik_y} & 0 & 2t'^z \cos k_z & t_1^x + t_2^x e^{-ik_x} \\ 0 & t_1^y + t_2^y e^{ik_y} & t_1^x + t_2^x e^{ik_x} & 2t^z \cos(k_z - \Phi) \end{bmatrix}. \quad (23)$$

**b** $n = 3$. The rotation symmetry operator is represented by

$$\hat{R}_{2\pi/3} = U_{R_{2\pi/3}} \mathcal{L}_{\frac{m G_z}{3}} \hat{r}_{2\pi/3}, \quad (24)$$

where $m = 1, 2$, and

$$U_{R_{2\pi/3}} = \begin{bmatrix} 0 & 0 & 1 & 0 & 0 & 0 \\ 0 & 0 & 0 & 1 & 0 & 0 \\ 0 & 0 & 0 & 0 & 1 & 0 \\ 0 & 0 & 0 & 0 & 0 & 1 \\ 1 & 0 & 0 & 0 & 0 & 0 \\ 0 & 1 & 0 & 0 & 0 & 0 \end{bmatrix}.$$

The Hamiltonian is given by

$$\mathcal{H}(\boldsymbol{k}) = \begin{bmatrix} 2t^z \cos(k_z) & t_1 & 0 & Je^{i\boldsymbol{k}\cdot\boldsymbol{a}_1} & 0 & t_2 \\ t_1 & 2t^z \cos(k_z - \Phi) & t_2 & 0 & Je^{i\boldsymbol{k}\cdot\boldsymbol{a}_2} & 0 \\ 0 & t_2 & 2t^z \cos(k_z - \Phi) & t_1 & 0 & Je^{i\boldsymbol{k}\cdot(-\boldsymbol{a}_1+\boldsymbol{a}_2)} \\ Je^{i\boldsymbol{k}\cdot\boldsymbol{a}_1} & 0 & t_1 & 2t^z \cos(k_z - 2\Phi) & t_2 & 0 \\ 0 & Je^{i\boldsymbol{k}\cdot\boldsymbol{a}_2} & 0 & t_2 & 2t^z \cos(k_z - 2\Phi) & t_1 \\ t_2 & 0 & Je^{i\boldsymbol{k}\cdot(-\boldsymbol{a}_1+\boldsymbol{a}_2)} & 0 & t_1 & 2t^z \cos(k_z) \end{bmatrix}, \quad (25)$$

where $\boldsymbol{a}_{1,2}$ are two lattice vectors in the $xoy$ plane.

**c** $n = 4$. The The rotation symmetry operator is represented by

$$\hat{R}_{\pi/2} = U_{R_{\pi/2}} \mathcal{L}_{\frac{m G_z}{4}} \hat{r}_{\pi/2}, \quad (26)$$

where m= $1, 2, 3$ and

$$U_{R_{\pi/2}} = \begin{bmatrix} 0 & 1 & 0 & 0 \\ 0 & 0 & 0 & 1 \\ 1 & 0 & 0 & 0 \\ 0 & 0 & 1 & 0 \end{bmatrix}.$$

The Hamiltonian is given by

$$\mathcal{H}(\boldsymbol{k}) = \begin{bmatrix} 2t^z \cos k_z & t_1 + t_2 e^{-ik_x} & t_1 + t_2 e^{-ik_y} & 0 \\ t_1 + t_2 e^{ik_x} & 2t^z \cos(k_z - \Phi) & 0 & t_1 + t_2 e^{-ik_y} \\ t_1 + t_2 e^{ik_y} & 0 & 2t^z \cos(k_z - 3\Phi) & t_1 + t_2 e^{-ik_x} \\ 0 & t_1 + t_2 e^{ik_y} & t_1 + t_2 e^{ik_x} & 2t^z \cos(k_z - 2\Phi) \end{bmatrix}. \quad (27)$$

**d** $n = 6$. The rotation symmetry operator is represented by

$$\hat{R}_{\pi/3} = U_{R_{\pi/3}} \mathcal{L}_{\frac{m G_z}{6}} \hat{r}_{\pi/3}, \quad (28)$$

where $m = 1, 2, 3, 4, 5$ and

$$U_{R_{\pi/3}} = \begin{bmatrix} 0 & 1 & 0 & 0 & 0 & 0 \\ 0 & 0 & 1 & 0 & 0 & 0 \\ 0 & 0 & 0 & 1 & 0 & 0 \\ 0 & 0 & 0 & 0 & 1 & 0 \\ 0 & 0 & 0 & 0 & 0 & 1 \\ 1 & 0 & 0 & 0 & 0 & 0 \end{bmatrix}.$$

The Hamiltonian is given by

$$\mathcal{H}(\boldsymbol{k}) = \begin{bmatrix} 2t^z \cos(k_z) & t & 0 & Je^{i\boldsymbol{k}\cdot\boldsymbol{a}_1} & 0 & t \\ t & 2t^z \cos(k_z - \Phi) & t & 0 & Je^{i\boldsymbol{k}\cdot\boldsymbol{a}_2} & 0 \\ 0 & t & 2t^z \cos(k_z - 2\Phi) & t & 0 & Je^{i\boldsymbol{k}\cdot(-\boldsymbol{a}_1+\boldsymbol{a}_2)} \\ Je^{i\boldsymbol{k}\cdot\boldsymbol{a}_1} & 0 & t & 2t^z \cos(k_z - 3\Phi) & t & 0 \\ 0 & Je^{i\boldsymbol{k}\cdot\boldsymbol{a}_2} & 0 & t & 2t^z \cos(k_z - 4\Phi) & t \\ t & 0 & Je^{i\boldsymbol{k}\cdot(-\boldsymbol{a}_1+\boldsymbol{a}_2)} & 0 & t & 2t^z \cos(k_z - 5\Phi) \end{bmatrix}. \quad (29)$$

## Supplementary Note 5. Numerical Demonstration of Stability

To demonstrate the stability of the topology with the addtion of atomic energy bands, we add an orbit in the middle of the unit cell as fig.3.a. The addition preserves the symmetry of our model since we put it on the wyckoff position of reflection. The atomic orbit has on site energy $\varepsilon'$ and couplings $\delta$ to other nearest orbits. The Hamiltonian now becomes

$$\bar{\mathcal{H}}(\boldsymbol{k}) = \begin{bmatrix} \varepsilon & [q_1^x(k_x)]^* & [q_+^y(k_y)]^* & 0 & \delta \\ q_1^x(k_x) & \varepsilon & 0 & [q_-^y(k_y)]^* & \delta \\ q_+^y(k_y) & 0 & -\varepsilon & [q_2^x(k_x)]^* & \delta \\ 0 & q_-^y(k_y) & q_2^x(k_x) & -\varepsilon & \delta \\ \delta & \delta & \delta & \delta & \varepsilon' \end{bmatrix}. \tag{30}$$

If we take $\varepsilon' = -2, \delta = 0$, and other parameters same with fig.4d in the main context, the atomic energy band merge into the valence band as Supplementary Fig.3.b. If we turn on the coupling strength $\delta$, we can see that the topological edge bands remain stable as fig3.c.

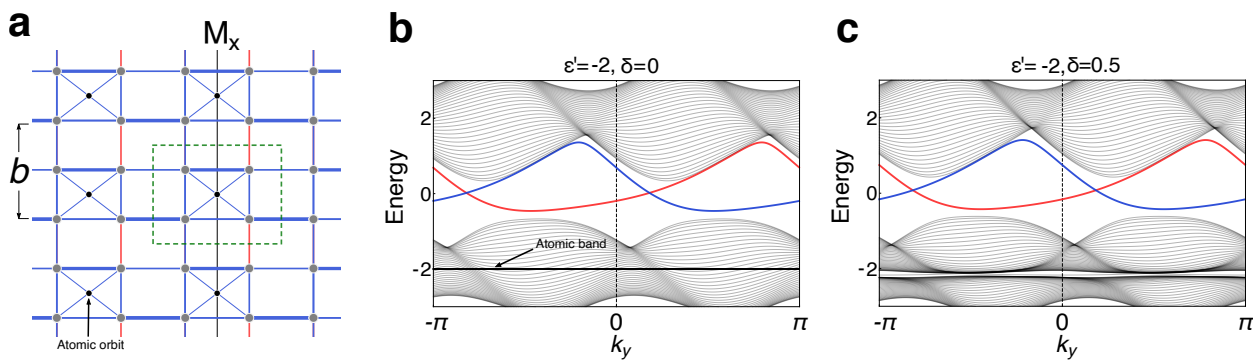

Supplementary Fig.3. **Stability of topology with adding of trivial bands**. **a** The addition of an atomic orbit in the middle of unit cell. **b** and **c** depict the corresponding band structures on a ribbon geometry with edges along y. We take the on site energy and coupling of the atomic band as $\varepsilon' = -2, \delta = 0$ and $\varepsilon' = -2, \delta = 0.5$ in **b** and **c** respectively.

## Supplementary Note 6. Gauge Fields in Physical Systems

In this section, we present a brief review on gauge fields in physical systems.

For condensed matter system, $\boldsymbol{Z}_2$ gauge fields can appear in quantum spin liquids. In the mean-field theory of quantum spin liquids, close to the ground states the spinors are coupled to gauge fields, for instance a $\boldsymbol{Z}_2$ gauge field (which defines $\boldsymbol{Z}_2$ spin liquids). In fact, probably this is the first time physicists noticed given gauge flux configurations can lead to projective representations of space groups (PSG), and thereby Xiao-Gang Wen proposed the PSG classification scheme for quantum phases of spin liquids.

We would like to point out a significant fact, i.e., $\boldsymbol{Z}_2$ gauge fields are essentially different from U(1) gauge fields in that time-reversal symmetry is preserved by $\boldsymbol{Z}_2$ gauge fields. This can be directly seen from the fact that real hopping amplitudes are sufficient to implement all $\boldsymbol{Z}_2$ gauge flux configurations, which naturally preserves time-reversal symmetry. More fundamentally, time reversal inverses flux, but $\pi = -\pi$ on a lattice. Thus, it is possible to realize $\boldsymbol{Z}_2$ gauge fields without introducing magnetic fields. While there could be other mechanisms, here we note the so-called dark-bright mechanism. The low-energy effective hopping amplitude of two sites through an intermediate high-energy site is negative $-t^2/\Delta$, with $\Delta$ the energy gap. This mechanism suggests that certain second-order hopping amplitudes can also take a negative phase in non-interacting crystalline systems.

In artificial systems, gauge fields can be engineered. This is an advantage for realizing our theoretical proposal. Below, we briefly introduce the mechanisms for generating gauge fields in these artificial systems.

- In photonic crystals, the artificial gauge field can be directly simulated by modulation of the resonant frequencies, i.e., adjusting the gap between site ring and link-ring wave guides. Other method is shaking the photonic crystal and the effective gauge field results from Floquet's theorem, similar to the cold atoms.

- In acoustic crystals, $\boldsymbol{Z}_2$ hopping phases can be readily realized by coupling the resonators with wave guides on different sides.

- In cold atom systems, the approaches of simulating the gauge field include rotating optical lattice, laser-assisted tunneling, and shaking the optical lattice. (i) Rotating optical lattice can only introduce the weak and uniform effective magnetic field and the side effect of Coriolis force should be compensated. (ii) For the laser-assisted tunneling, the atomic hopping with desired gauge potentials are engineered by coupling internal levels of atoms with laser beams and thus different kinds of gauge fields can be simulated even including the nonabelian one. (iii) The advantage of shaking optical lattice relies on that it is free of internal levels.

- In mechanical systems, effective $\boldsymbol{Z}_2$ gauge field can be simulated by just tuning the stiffness coefficients of the springs.

- For the electric circuits, the $\boldsymbol{Z}_2$ gauge field can be realized by tuning the capacitances and inductances.



\end{document}